# An Improved Decentralized Control of Grid-Connected Cascaded Inverters with Different Power Capacities

Xiaochao Hou, Yao Sun, Xin Zhang, Jinsong He, and Josep Pou, *Fellow, IEEE*

*Abstract*—The existing decentralized control for cascaded inverters is based on the assumption that all modules have same capacities, and a local fixed-amplitude-varied-phase voltage control is adopted for each inverter module. However, available source power capacities of cascaded inverters may be different in some practical applications. To address this issue, this letter proposes an improved decentralized control scheme, in which the voltage amplitudes are varied according to their individual available powers. Moreover, a power factor consistency control is proposed to achieve autonomous voltage phase synchronization. The steady-state analysis and synchronization mechanism of cascaded inverters are illustrated. In addition, the proposed strategy has other advantages, such as, adjustable grid power factor and immune to the grid voltage fault. The effectiveness of the proposed control is tested by experiments.

## I. INTRODUCTION

Cascaded H-bridge converters are widely applied in the fully modular inverter system [1]-[2] and cascaded distributed generation system [3]-[6]. Specifically, the cascaded inverters have two operation modes: an islanded mode for feeding a load and a grid-connected mode for connecting to utility grid.

In the islanded mode, the decentralized control have been gradually studied for modular converters due to the advantages of full modularity, communication-free and high reliability. In [7], a compound decentralized control strategy is early proposed for input-series-output-series DC/DC converters to maintain autonomous voltage and power balances. Then, some improved decentralized control methods are proposed to eliminate the input-voltage sensors [8] and enhance the dynamic voltage regulation [9]. While for the cascaded DC/AC inverters, literature [10] firstly uses the frequency self-synchronization control to achieve the decentralized power balance. However, the system stability of islanded operation highly relies on the load characteristic and only the resistive-inductive load is applicable. To overcome the limitation, an adaptive droop control is proposed to adapt arbitrary resistive-inductive-capacitive load in [11]. Nevertheless, the aforementioned [10]-[11] only focused on the islanded mode, which cannot be directly adopted to the grid-connected mode.

In the grid-connected mode, the decentralized control is rarely reported and need more concern. In the latest research, literature [12] presents a fully decentralized control for cascaded inverters. As all modules have the fixed-voltage-amplitude and the same flowing-current, their apparent powers are equal all the time. As a result, it is just suitable for modular inverters with same capacities and same power outputs [12]. However, the output powers of these inverters are required to be different for flexible power management in some special applications, such as in cascaded PV micro-inverters [4]-[5] and cascaded storage inverters [6]. In these cascaded distributed generation (DG) systems [3]-[6], the decentralized control of [12] cannot provide the independent power-regulation function.

To solve this limitation of [12], this letter proposes a novel decentralized control for grid-connected cascaded inverters by considering the different power capacities. Compared to [12], the proposed method has the following advantages:

- *Independent power-regulation for each inverter*. This study adopts a varied-amplitude-fixed-phase voltage control to independently regulate the power output of each inverter, which is more flexible than the fixed-amplitude-varied-phase voltage control of [12].
- *Arbitrarily adjustable grid power factor*. In [12], there is a tradeoff between grid power factor (PF) and stability margin, and unity PF cannot be realized. But in this study, arbitrary PF can be realized without compromising the system stability.
- *Adapt grid voltage fluctuation*. In [12], the system stability is sensitive to grid voltage sag. Instead, the proposed control can adapt to the grid voltage fluctuation.
- *More generalized and practical*. Not only fully modular inverters but also asymmetrical cascaded inverters are applicable. Each inverter is controlled independently without communicating with other inverters, resulting in a high reliability and scalability.

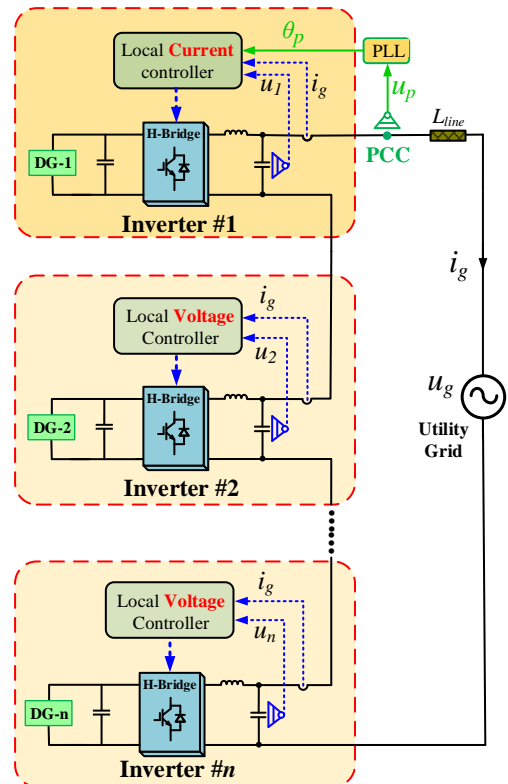

Fig. 1. Configuration of grid-connected cascaded inverters.

## II. IMPROVED DECENTRALIZED CONTROL

### A. Structure of Grid-Connected Cascaded Inverters

Fig. 1 shows the system structure and overall control framework of grid-connected cascaded inverters. The system consists of *n* cascaded DG-based inverters. From Fig.1, some main features are clarified that:

- *Individual local controllers for inverters.* Each inverter is controlled by a local controller, which needs only local information.
- *Common grid flowing current.* All inverters have the common grid flowing current. Thus, the fundamental-frequency current component can be used as an inherent synchronizer, and the synchronization of voltage phase can be realized by power-factor-angle consistency.
- *Current-controlled for inverter-1.* To guarantee a required grid power factor (PF), inverter-*1* is controlled as a current source, regulating the grid flowing current, which need to acquire the voltage phase from the point of common coupling (PCC).
- *Voltage-controlled for remaining n-1 inverters.* The rest of inverters are controlled as voltage sources to ensure independent power-regulation and self-synchronization.

### B. Improved Decentralized Control

Fig. 2 presents the current-controlled scheme of inverter-*1*. The grid current reference $i_g^*$ is generated by synthesizing the current amplitude $I_g$ and phase $\theta_{Ig}$, which are determined by

$$\begin{cases} I_g = \left(K_{Pi} + \dfrac{K_{Ii}}{s}\right)(P_i^* - P_i) \\ \theta_{Ig} = \theta_p - \varphi^* \end{cases} \quad (i=1) \quad (1)$$

where $P_1^*$ denotes the maximum power capacity from the primary DG-*1* source. $P_1$ is output active power of inverter-*1*. $K_{Pi}$ and $K_{Ii}$ are the proportional-integral (PI) coefficients. $\theta_p$ is the voltage phase of PCC. $\varphi^*$ is the predesigned PF angle, which can be flexibly set by considering the grid requirement and reactive-power compensation capability of each inverter. Particularly for cascaded PV micro-inverters, $P_i^*$ can be determined by the maximum power point tracking (MPPT) algorithm, and a unity PF can be realized by setting $\varphi^*=0$.

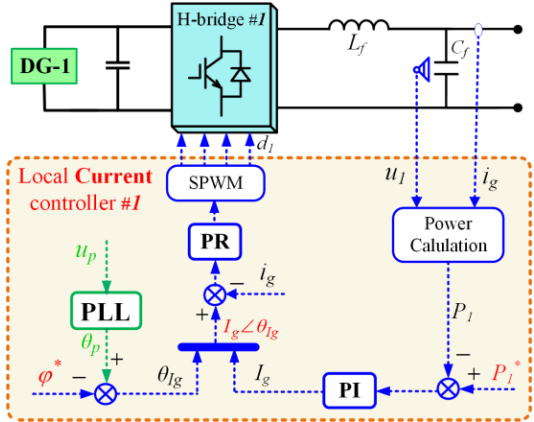

Fig. 2. Current-controlled scheme of inverter-*1*.

Fig. 3 presents the voltage-controlled scheme of inverter-*i*. The output voltage reference $u_i^*$ is generated by synthesizing the voltage amplitude $V_i$ and phase $\theta_i$, which are determined by an active power-voltage amplitude (*P-V*) control and a power factor-frequency ($\varphi$-$\omega$) control, respectively.

$$\begin{cases} V_i = \dfrac{V_g^*}{n} + \left(K_{Pvi} + \dfrac{K_{Ivi}}{s}\right)(P_i^* - P_i) \\ \dot{\theta}_i = \omega_i = \omega^* + \left(K_{P\omega i} + \dfrac{K_{I\omega i}}{s}\right)(\varphi^* - \varphi_i) \end{cases} \quad (i=2,3,...,n) \quad (2)$$

where $P_i^*$ denotes the maximum power capacity from the primary DG-*i* source. $P_i$ is output active power of inverter-*i*. $K_{Pvi}$ and $K_{Ivi}$ are the PI coefficients of amplitude control. $V_g^*$ denotes the nominal grid voltage amplitude. $\omega^*$ denotes the nominal grid angular frequency. $K_{P\omega i}$ and $K_{I\omega i}$ are the PI coefficients of frequency control. $\varphi^*$ is the predesigned PF angle. $\varphi_i$ is output PF angle, which is calculated by the difference of output voltage phase and current phase.

$$\varphi_i = \theta_i - \theta_{Ig} \quad (i=2,3,...,n) \quad (3)$$

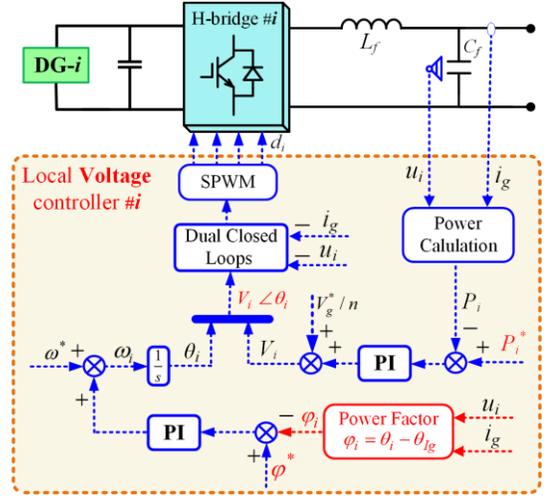

Fig. 3. Voltage-controlled scheme of inverter-*i*. ($i=2,3,...,n$)

### C. Steady-State Analysis

Fig. 4 shows the voltage phasor diagram in steady state. Due to the zero steady-state error of PI control in (2), the PF angles of inverters would be identical ($\varphi_2=\varphi_3=...=\varphi_n=\varphi^*$). Then, by combining (1)-(3), the same voltage phases are obtained because of the common grid current

$$\theta_p = \theta_1 = \theta_2 = \cdots = \theta_n = (\theta_{Ig} + \varphi^*) \quad (4)$$

Due to the same PF angles and grid current of all inverters, the active power flow $P_i^*=V_iI_g\cos\varphi^*$ is in proportion to the voltage amplitude $V_i$ in steady-state.

$$P_1^* : P_2^* : \cdots : P_n^* = V_1 : V_2 : \cdots : V_n \quad (5)$$

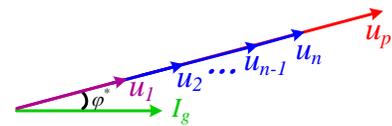

Fig. 4. Voltage phasor diagram in steady-state.

### D. Synchronization-Mechanism Analysis

As the grid current measured by all inverters is the same, the fundamental current component is used as a common synchronization baseline. That is, the voltage phase synchronization of inverters can be realized by power-factor-angle consistency.

To better understand the proposed PF control, the synchronization mechanism analysis is carried out. Fig. 5 presents the equivalent circuit and phasor diagrams of grid-connected cascaded inverters. For simplicity, we assume a unity PF by setting $\varphi^*=0$.

In Fig. 5(b), $u_i$ (*blue phasor*) leads the steady-state grid current (*green phasor*); while $u_j$ (*red phasor*) lags it. Initially, $\varphi_i>0$, $\varphi_j<0$. Then, $\omega_i<\omega^*<\omega_j$ is obtained from (2). As a result, $\varphi_i$ decreases ($\Delta\omega_i=\omega_i-\omega^*<0$), while $\varphi_j$ increases ($\Delta\omega_j=\omega_j-\omega^*>0$). The convergence process will continue until $\varphi_i=\varphi_j=\varphi^*=0$, and $\theta_i=\theta_j=(\theta_{Ig}+\varphi^*)$.

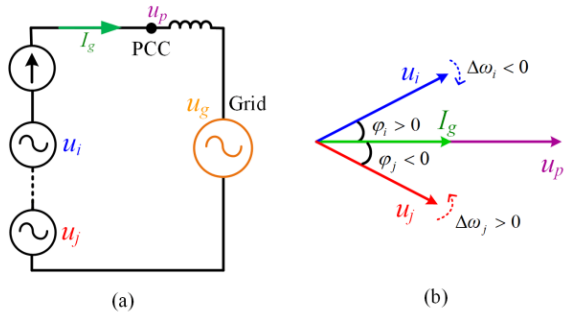

Fig. 5. A cascaded inverters system for synchronization-mechanism analysis. (a) Equivalent circuit. (b) Phasor diagram.

## III. EXPERIMENT RESULTS

To verify the feasibility of the proposed control, a grid-connected system comprised of three cascaded inverters has been built and tested in the lab. The system parameters are listed in Table I. The switching frequency of inverters is 10 kHz. The front-end DC-link of inverters is an ideal DC constant-voltage source. The nominal grid voltage is 311V/50 Hz.

TABLE I
EXPERIMENT PARAMETERS

| Symbol | Value | Symbol | Value |
|---|---|---|---|
| $V_g^*$ | 311 V | $n$ | 3 |
| $\omega^*$ | 100π rad/s | $K_{Pvi}$ | 0.2 |
| $L_{line}$ | 0.3mH | $K_{Ivi}$ | 0.6 |
| $\varphi^*$ | 0 (case-1) | $K_{P\omega i}$ | 2 |
| | 0.128π (case-2) | $K_{I\omega i}$ | 0.2 |

### A. Case-1: Source Power Change under Unity PF

To evaluate the independent power-regulation capability of each inverter, the condition of different DG-source power is considered in case-1. As experiment results shown in Fig. 6, the available powers of three inverters are changed from same values to different values at t=1s.

Before t=1s, output voltages $u_1$, $u_2$, $u_3$ of three inverters are identical in Fig. 6(b), and output active-powers $P_1$, $P_2$, $P_3$ of three inverters are equal to 1.5kW in Fig. 6(c). After t=1s, $P_2$ changes from 1.5kW to 1.3kW, and $P_3$ changes to 1.1kW while $P_1$ is unchanged. From the steady-state voltage in Fig. 6(a), output voltages $u_1$, $u_2$, $u_3$ have the same phase with the grid current $i_g$, which reveals that a predesigned unity PF is realized. Meanwhile, the voltage amplitude $V_i$ is proportion to the output active power $P_i^*$ in steady-state, which verifies the feasibility of the proposed varied-amplitude-fixed-phase voltage control. From case-1, the proposed method can work in the different source power conditions and achieve unity PF.

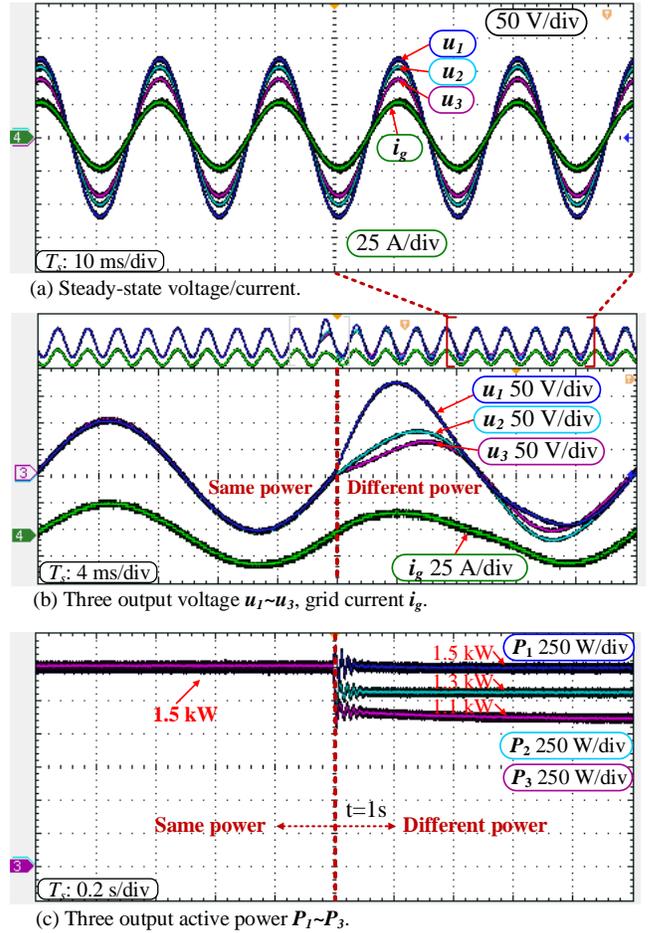

Fig. 6. Experiment results of case-1.

### B. Case-2: Grid Voltage Sag under No-Unity PF

Different from case-1, the testing of case-2 adopts a no-unity PF ($\cos\varphi^*=0.92$) to present the PF controllability of the proposed strategy. In this condition, each inverter provides reactive-power compensation for utility grid. Moreover, to demonstrate the effectiveness of the proposed strategy under grid contingencies, a 15% grid voltage sag is imposed at t=1s.

As shown in Fig. 7, before t=1s, output voltages $u_1$, $u_2$, $u_3$ of three inverters are in same phase, and the voltage amplitude is proportional to the active power outputs, where $P_1$=1.5kW, $P_2$=1.3kW, $P_3$=1.1kW. After t=1s, to compensate for the grid voltage sag, output voltages $u_1$, $u_2$, $u_3$ react immediately in Fig. 7(a). After about two cycles, $u_1$, $u_2$, $u_3$ reach the new steady states, and the grid current amplitude $I_g$ becomes higher to guarantee the unchanged active power output. As shown in Fig. 7(b)-(c), the active/reactive powers have a satisfactory dynamic response.

Moreover, from the active/reactive power values in steady-state, the same PF=0.92 is derived for three inverters. Thus, a predesigned no-unity PF is realized to provide a reactive-power compensation. From case-2, the proposed method can adapt to the grid voltage fluctuation and achieve a flexible PF regulation.

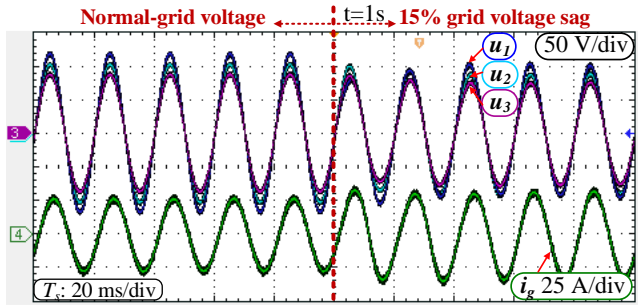

(a) Three output voltage $u_1 \sim u_3$, grid current $i_g$.

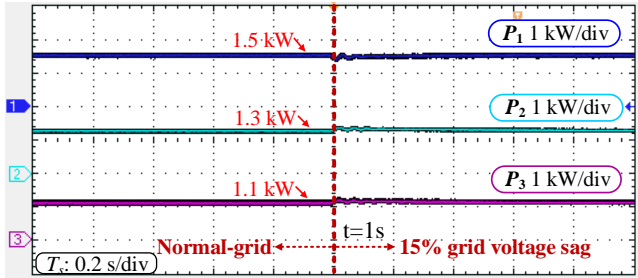

(b) Three output active power $P_1 \sim P_3$.

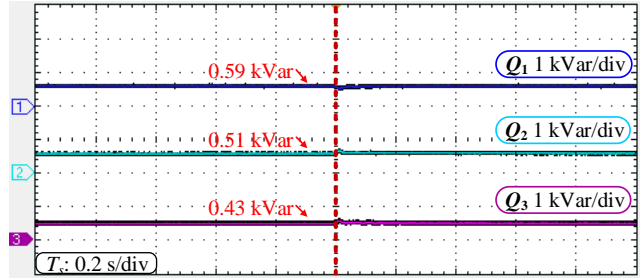

(c) Three output reactive power $Q_1 \sim Q_3$.

Fig. 7. Experiment results of case-2.

## IV. CONCLUSION

This letter proposes an improved decentralized control of grid-connected cascaded inverters. Independent power-regulation for each inverter can be obtained by the proposed varied-amplitude-fixed-phase voltage control. The voltage amplitude is varied according to the primary source power. The voltage-phase synchronization is achieved by the power-factor-angle consistency. Compared with the existing fixed-amplitude-varied-phase method, the proposed strategy has three main advantages: 1) suitable for asymmetrical cascaded DG sources; 2) adjustable grid power factor; 3) immune to the grid voltage fault. As only local information is necessary for each inverter, the proposed method has the advantages of high reliability and scalability, which has promising applications in large-scale cascaded PV-based and storage-based inverters.

## V. REFERENCES

[1] T. Fang, X. Ruan and C. K. Tse, "Control Strategy to Achieve Input and Output Voltage Sharing for Input-Series–Output-Series-Connected Inverter Systems," *IEEE Trans. Power Electron.*, vol. 25, no. 6, pp. 1585-1596, Jun. 2010.

[2] D. Sha, G. Xu and X. Liao, "Control strategy for input-series-output-series high-frequency AC-link inverters," *IEEE Trans. Power Electron.*, vol. 28, no. 11, pp. 5283-5292, Nov. 2013.

[3] A. Mortezaei, M. G. Simoes, A.S. Bubshait, T.D.C. Busarello, F.P. Marafao, A. Al-Durra, "Multifunctional control strategy for asymmetrical cascaded H-bridge inverter in microgrid applications," *IEEE Trans. Ind. Appl.*, pp. 1–14, Nov. 2016.

[4] J. He, Y. Li, C. Wang, Y. Pan, C. Zhang and X. Xing, "Hybrid Microgrid With Parallel- and Series-Connected Microconverters," *IEEE Trans. Power Electron.*, vol. 33, no. 6, pp. 4817-4831, Jun. 2018.

[5] H. D. Tafti, A. I. Maswood, G. Konstantinou, C. D. Townsend, P. Acuna and J. Pou, "Flexible Control of Photovoltaic Grid-Connected Cascaded H-Bridge Converters During Unbalanced Voltage Sags," *IEEE Trans. Ind. Electron.*, vol. 65, no. 8, pp. 6229-6238, Aug. 2018.

[6] G. Wang, G. Konstantinou, C. D. Townsend, J. Pou, S. Vazquez, G. D. Demetriades, V. G. Agelidis, "A Review of Power Electronics for Grid Connection of Utility-Scale Battery Energy Storage Systems," *IEEE Trans. Sustainable Energy*, vol. 7, no. 4, pp. 1778-1790, Oct. 2016.

[7] W. Chen, X. Ruan, H. Yan and C. K. Tse, "DC/DC conversion systems consisting of multiple converter modules: stability, control, and experimental verifications," *IEEE Trans. Power Electron.*, vol. 24, no. 6, pp. 1463-1474, Jun. 2009.

[8] D. Sha, Z. Guo, T. Luo and X. Liao, "A General Control Strategy for Input-Series–Output-Series Modular DC–DC Converters," *IEEE Trans. Power Electron.*, vol. 29, no. 7, pp. 3766-3775, Jul. 2014.

[9] W. Chen and G. Wang, "Decentralized Voltage-Sharing Control Strategy for Fully Modular Input-Series–Output-Series System With Improved Voltage Regulation," *IEEE Trans. Power Electron.*, vol. 62, no. 5, pp. 2777-2787, May 2015.

[10] J. He, Y. Li, B. Liang and C. Wang, "Inverse power factor droop Control for decentralized power sharing in series-connected-microconverters-based islanding microgrids," *IEEE Trans. Ind. Electron.*, vol. 64, no. 9, pp. 7444-7454, Sep. 2017.

[11] Y. Sun, G. Shi, X. Li, W. Yuan, M. Su, H. Han, and X. Hou, "An f-P/Q droop control in cascaded-type microgrid, " *IEEE Trans. Power Systems*, vol. 33, no. 1, pp. 1136-1138, Jan. 2018.

[12] X. Hou, Y. Sun, H. Han, Z. Liu, W. Yuan and M. Su, "A Fully Decentralized Control of Grid-Connected Cascaded Inverters," *IEEE Trans. Sustainable Energy*, vol. 10, no. 1, pp. 315-317, Jan. 2019.